\documentclass[useAMS,usenatbib]{mn2e}

\usepackage{aecompl}

\usepackage{graphicx}
\usepackage{amssymb}
\usepackage{amsmath}
\usepackage{aas_macros}
\usepackage{deluxetable}
\usepackage{lscape}
\usepackage{hyperref}
\usepackage{rotating}
\usepackage[symbol*]{footmisc}



\title[Investigating the Global Collapse of Filaments Using Smoothed Particle Hydrodynamics]{Investigating the Global Collapse of Filaments Using Smoothed Particle Hydrodynamics}
\author[S. D. Clarke and A. P. Whitworth]{S. D. Clarke$^{1}$\thanks{E-mail: seamus.clarke@astro.cf.ac.uk } and A. P. Whitworth$^{1}$ \\$^{1}$School of Physics and Astronomy, Cardiff University, Cardiff, CF24 3AA, UK}

\newcommand{\OO}{_{_{\rm O}}}
\begin{document}

\date{}

\pagerange{\pageref{firstpage}--\pageref{lastpage}} \pubyear{2002}

\maketitle

\label{firstpage}

\begin{abstract}
We use Smoothed Particle Hydrodynamic simulations of cold, uniform density, self-gravitating filaments, to investigate their longitudinal collapse timescales; these timescales are important because they determine the time available for a filament to fragment into cores. A filament is initially characterised by its line-mass, $\mu\OO$, its radius, $R\OO$ (or equivalently its density $\rho\OO\!=\!\mu\OO/\pi R\OO^2$), and its aspect ratio, $A\OO\;\,(\equiv Z\OO/R\OO$, where $Z\OO$ is its half-length). The gas is only allowed to contract longitudinally, i.e. parallel to the symmetry axis of the filament (the $z$-axis). \citet{Pon12} have considered the global dynamics of such filaments analytically. They conclude that short filaments ($A\OO\!\la\!5$) collapse along the $z$-axis more-or-less homologously, on a time-scale $t_{_{\rm HOM}} \sim 0.44\,A\OO\,(G\rho\OO)^{-1/2}$; in contrast, longer filaments ($A\OO\!\ga\!5$) undergo end-dominated collapse, i.e. two dense clumps form at the ends of the filament and  converge on the centre sweeping up mass as they go, on a time-scale $t_{_{\rm END}} \sim 0.98\,A\OO^{1/2}\,(G\rho\OO)^{-1/2}$. Our simulations do not corroborate these predictions. First, for all $A\OO\!\ga\!2$, the collapse time satisfies a single equation
\[t_{_{\rm COL}}\;\sim\;(0.49+0.26A\OO)(G\rho\OO)^{-1/2}\,,\]
which for large $A\OO$ is much longer than the Pon et al. prediction. Second, for all $A\OO\!\ga\!2$, the collapse is end-dominated. Third, before being swept up, the gas immediately ahead of an end-clump is actually accelerated outwards by the gravitational attraction of the approaching clump, resulting in a significant ram pressure. For high aspect ratio filaments the end-clumps approach an asymptotic inward speed, due to the fact that they are doing work both accelerating and compressing the gas they sweep up. Pon et al. appear to have neglected the outward acceleration and its consequences.
\end{abstract}


\section{Introduction}%

The ubiquity of non-spherical structures in molecular clouds \citep[e.g.][]{SchElm97,Lad99,Mye09,And10} has lead to many papers on the effects of dimensionality on global collapse \citep[e.g.][]{Bas83,Bas91,BurHar04,Pon11,Toa12,Pon12}. It has been shown analytically that non-spherical structures collapse on longer timescales than spherical clouds of the same density \citep[e.g.][]{BurHar04,Pon12}; and this has been corroborated by simulations \citep[e.g.][]{Bas83,Vaz07}. In cylindrical clouds -- hereafter filaments -- there are two possible longitudinal collapse modes (i.e. collapse modes parallel to the symmetry axis of the filament): a filament may collapse approximately homologously (hereafter homologous collapse), or clumps may form at the ends and move inwards sweeping up material as they go \citep[hereafter end-dominated collapse;][]{Bas83}.  

\begin{figure*}
  \centering
  \includegraphics[width = 0.49\linewidth]{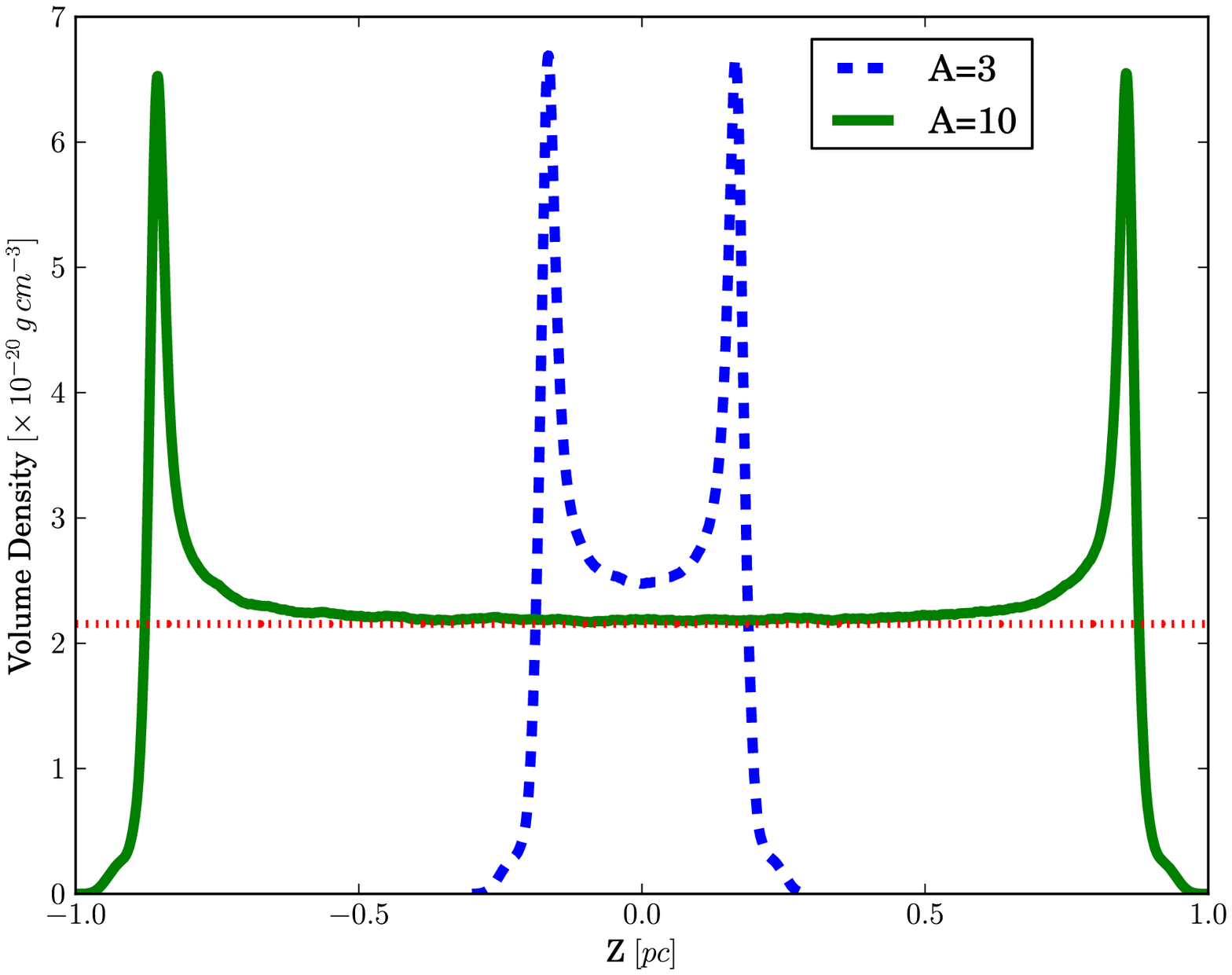}
  \includegraphics[width = 0.49\linewidth]{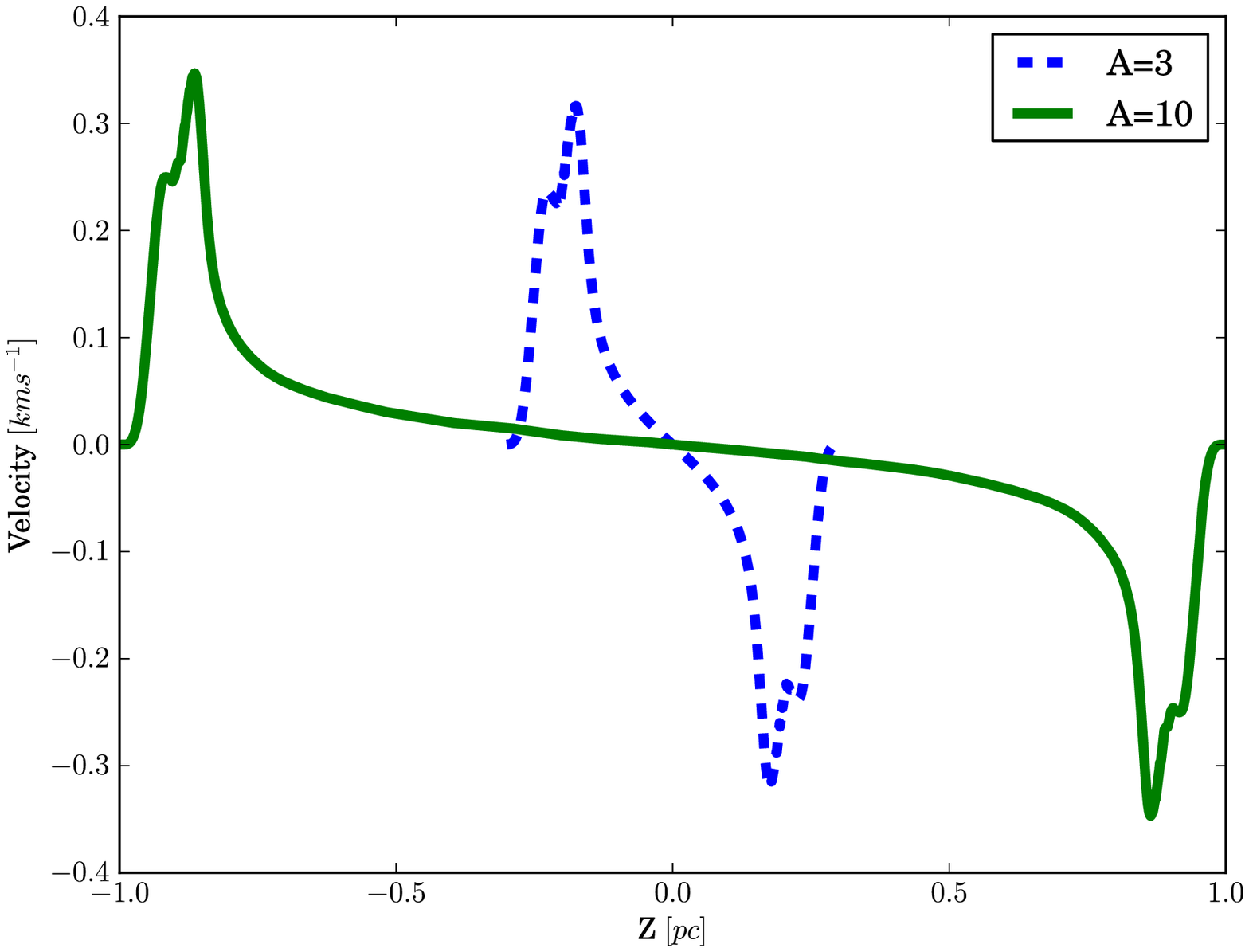}
  \caption{(a) The volume-density profile, and (b) the velocity profile, along the $z$-axis at time $t=0.55\,\rm{Myr}$, from the simulations of filaments having $\mu\OO=10\, {\rm M}_{_\odot}\,\rm{pc}^{-1}$, $R\OO=0.1\,{\rm pc}$ and $A\OO=3$ (dashed curves) or $A\OO=10$ (solid curves). The signatures of end-dominated collapse are clearly seen in both, viz. strong density peaks and supersonic inwards motions at the extremes, representing the end-clumps that dominate the dynamics. The horizontal dotted line in figure (a) is the initial volume density; interior material is relatively unaffected by the collapse until the density peaks reach it.}
\label{fig:veldenprofile10}
\end{figure*}

Recently, \citet{Pon12} have explored analytically the effect that these two modes have on the freefall collapse time for a uniform filament. During homologous collapse the density of a filament remains approximately uniform but changes with time, so all points within the filament have approximately the same collapse timescale. \citet{Pon12} calculate that the homologous collapse timescale is
\begin{equation}
\label{eq:homocollapse}
t_{_{\rm HOM}}\;\,\simeq\;\,0.44\,A\OO\,\left(G\rho\OO\right)^{-1/2}\, .
\end{equation}
Where $\rho\OO$ is the initial volume density of the filament, $A\OO$ is the initial aspect ratio defined as $A\OO\;\,\equiv Z\OO/R\OO$,  $Z\OO$ and $R\OO$ are the filament's half-length and radius respectively.

The exact equation describing the longitudinal acceleration inside a filament is non-linear, and the non-linear terms become very important at the ends of a filament. Consequently the ends of the filament are accelerated to much greater speeds than in homologous collapse, causing them to sweep up interior matter as they overtake it. \citet{Pon12} calculate that in this case the collapse time is
\begin{equation}
\label{eq:edgecollapse}
t_{_{\rm END}}\;\,\simeq\;\,0.98\,A\OO^{1/2}\,\left(G\rho\OO\right)^{-1/2}\,.
\end{equation}

Since the collapse times for the two modes depend on $A\OO$ in different ways, \citet{Pon12} conclude that filaments collapse on whichever is the shorter timescale, i.e. short filaments collapse approximately homologously, on a time-scale given by Eqn. (\ref{eq:homocollapse}), and long ones collapse via the end-dominated mode, on a timescale given by Eqn. (\ref{eq:edgecollapse}). Comparing Eqns. (\ref{eq:homocollapse}) and (\ref{eq:edgecollapse}), they infer that the switch occurs at $A\OO \sim 5$.

The two modes should be distinguishable by their kinematics. A short filament collapsing homologously has an approximately linear velocity profile along its long axis, while a long filament collapsing via the end-mode has low inward velocities except at its ends, where the inward velocity increases dramatically. 

In this paper we present numerical simulations of initially uniform filaments to evaluate their freefall time-scales, and to investigate the collapse-mode dichotomy presented in \citet{Pon12}. In \S \ref{SEC:NUM} we detail the simulation setup and the initial conditions used; in \S \ref{SEC:RES} we present the results of these simulations; in \S \ref{SEC:DIS} we discuss their significance and support the discussion with a semi-analytical model of the system dynamics; and in \S \ref{SEC:CON} we summarise our conclusions. 

\begin{figure}
\centering
\includegraphics[width = 0.98\linewidth]{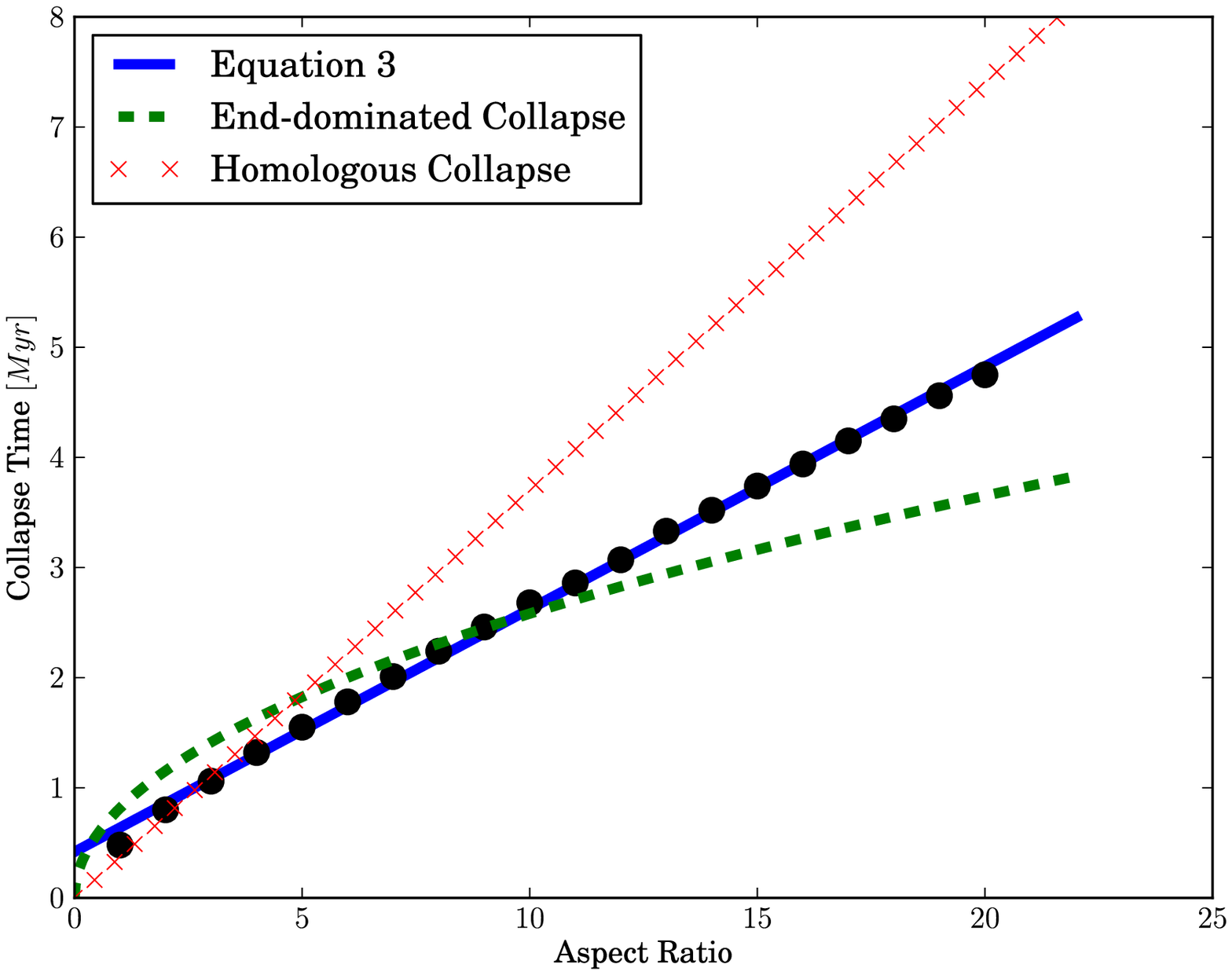}
\includegraphics[clip,width = 0.98\linewidth]{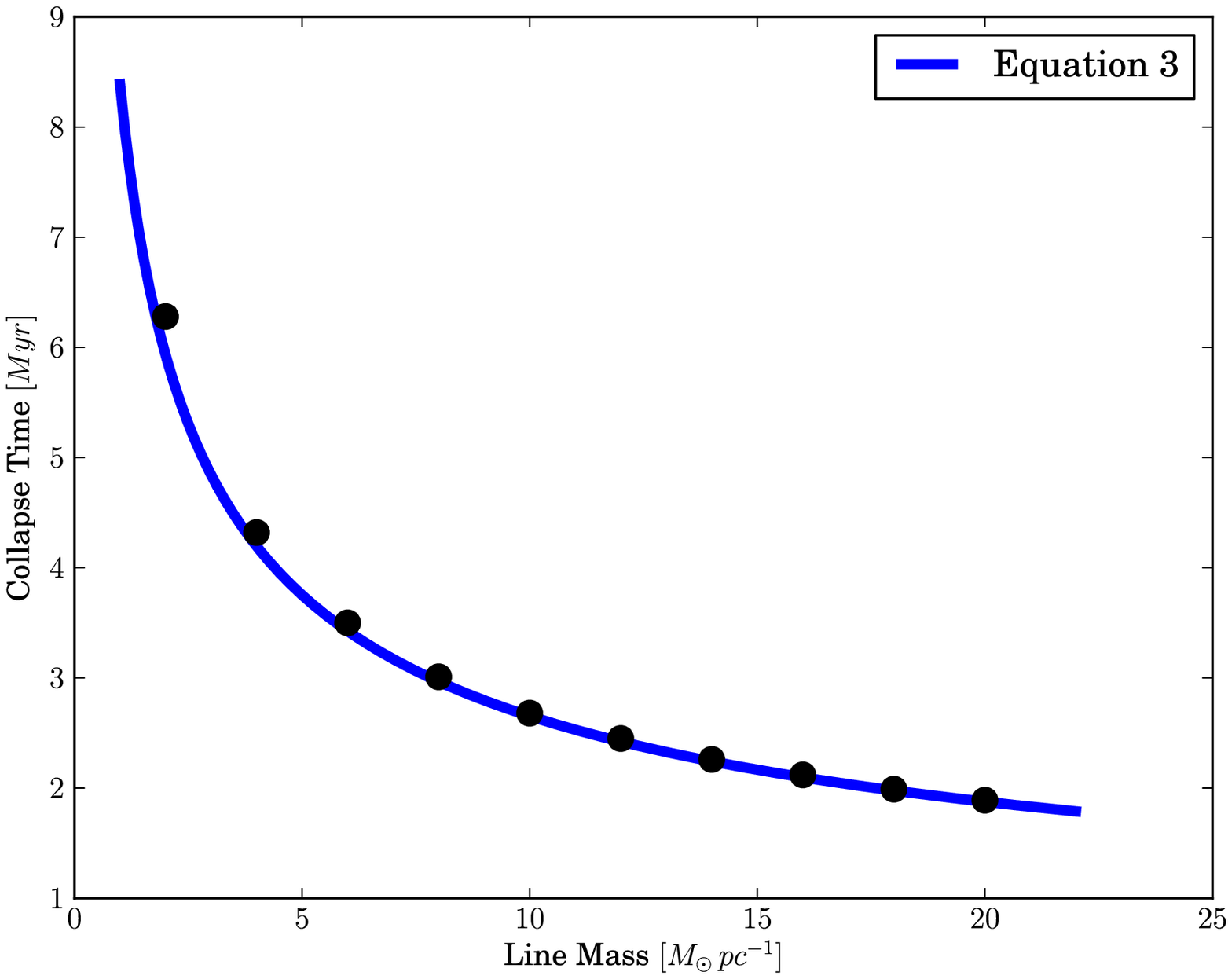}
\includegraphics[clip,width = 0.98\linewidth]{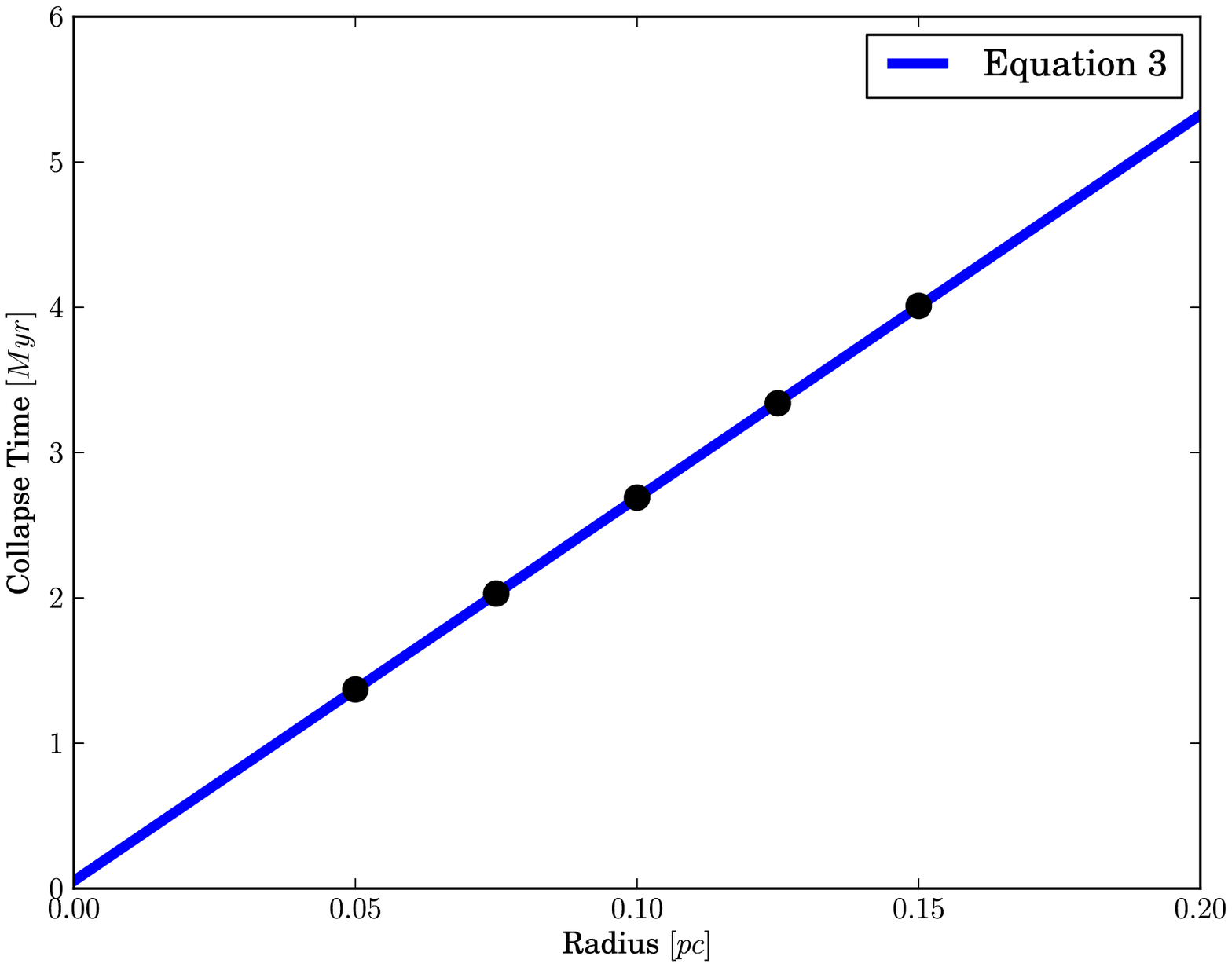}
\caption{The filled circles show values of the collapse time, $t_{_{\rm COL}}$, obtained when (a) the aspect ratio, $A\OO$, is varied and $\mu\OO$ and $R\OO$ are fixed at their fiducial values; (b) the line-density, $\mu\OO$, is varied and $A\OO$ and $R\OO$ are fixed at their high aspect ratio fiducial values; (c) the radius, $R\OO$, is varied and $A\OO$ and $\mu\OO$ are fixed at their high aspect ratio fiducial values. The high aspect ratio fiducial values are $A\OO\!=\!10$, $\mu\OO\!=\!10\,{\rm M}_{_\odot}\,{\rm pc}^{-1}$, and $R\OO\!=\!0.1\,{\rm pc}$. The blue continuous lines are the predictions of Eqn. (\ref{eq:collapsetime}). The red dotted and green dashed lines are the predictions from \citet{Pon12} (Eqns. \ref{eq:homocollapse} and \ref{eq:edgecollapse} respectively) }
\label{fig:varythings}
\end{figure}

\begin{figure}
\centering
\includegraphics[width = 0.98\linewidth]{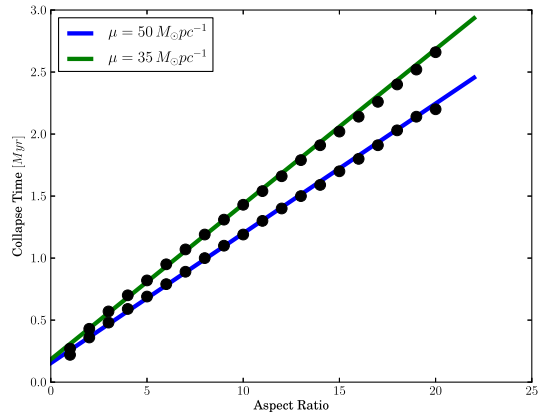}
\caption{The filled circles show values of the collapse time, $t_{_{\rm COL}}$, obtained with different aspect ratios, $A\OO$, when the radius, $R\OO$, is fixed at $0.1\,{\rm pc}$, and the line-density is fixed at $\mu\OO=35\,{\rm M}_{_\odot}\,{\rm pc}^{-1}$ or $\mu\OO=50\,{\rm M}_{_\odot}\,{\rm pc}^{-1}$. The continuous lines are the predictions of Eqn. \ref{eq:collapsetime} with $\rho\OO=\mu\OO/\pi R\OO^2$.}
\label{fig:predict}
\end{figure}

\section{Numerical Setup}\label{SEC:NUM}%

The simulations presented in this paper are performed using the Smoothed Particle Hydrodynamics (SPH) code GANDALF (Hubber et al. in prep). The simulations invoke both gravity and hydrodynamics, with an isothermal equation of state. Grad-h SPH \citep{PriMon04} is implemented, with $\eta = 1.2$ \footnote[2]{$\eta$ is a constant used to calculate smoothing lengths self-consistently in Grad-h SPH \citep[see e.g.][]{Hub11}. It determines the average number of neighbours.}, so that a typical particle has $\sim 56$ neighbours. Although we are interested in freefall collapse, the gas is given a non-zero temperature, $T=1\,{\rm K}$, in order to capture shocks, and to stop artificial clumping of SPH particles. For molecular gas with solar composition, this results in an isothermal sound speed of $0.07 \; \rm km \, \rm s^{-1}$. Running identical simulations at different low temperatures produces only small variations in the collapse times, so we are satisfied that $T=1\,\rm{K}$ gives a reasonable approximation to free-fall.

The code is modified so that radial motions are suppressed, i.e. $v_{x} = v_{y} = 0$ for all particles. Thus, the longitudinal collapse of a cylindrical cloud with super-critical line mass can be simulated without it collapsing radially. Such a constraint has no effect on the longitudinal motion of the gas and allows the simulations to conform to the constant radius condition adopted by \citet{Pon12}. 

A hexagonal close-packed grid of particles is used to produce a uniform density. The grid is rotated around the 3 axes by random angles to ensure that the grid's symmetry axes are not aligned with the Cartesian axes, and a cylindrical filament is then cut from this grid. The initial conditions are characterised by \\
\indent $A\OO$, the aspect ratio of the filament; \\
\indent $\mu\OO$, the mass per unit length of the filament; and \\
\indent $R\OO$, the radius of the filament. \\
$A\OO$ is varied between $1$ and $20$. A filament is defined as a structure with an aspect ratio of at least 3 \citep[e.g.][]{Pan14}; aspect ratios below this value are only simulated for completeness. $\mu\OO$ is varied between $2\,\rm M_{_\odot}\,\rm{pc}^{-1}$ and $50\,\rm M_{_\odot}\,\rm{pc}^{-1}$. $R\OO$ is varied between $0.05\,{\rm pc}$ and $0.15\,{\rm pc}$. This results in a range of initial number densities, from $ 5 \, \times \, 10^{2} \; \rm cm^{-3} $ to $ 1 \, \times \, 10^{5} \; \rm cm^{-3} $, for molecular gas of solar composition. 

Much of our discussion will consider two fiducial cases: a high aspect ratio case characterised by $A\OO\!=\!10$, $\mu\OO\!=\!10\,{\rm M}_{_\odot}\,{\rm pc}^{-1}$ and $R\OO\!=\!0.1\,{\rm pc}$ and a low aspect ratio case characterised by $A\OO\!=\!3$, $\mu\OO\!=\!10\,{\rm M}_{_\odot}\,{\rm pc}^{-1}$ and $R\OO\!=\!0.1\,{\rm pc}$.

The simulations are all run with the same initial volume-density of SPH particles, $n_{_{\rm SPH}}\!\simeq\! 800,000 \; \rm pc^{-3}$, to ensure that they have comparable spatial resolution, $\sim 0.01\,{\rm pc}$; the mass resolution therefore varies between $0.008\,{\rm M}_{_\odot}$ and $0.20\,{\rm M}_{_\odot}$. Having comparable spatial resolution (rather than comparable mass resolution) is appropriate since we are concerned here with the longitudinal density- and velocity-profiles (rather than fragmentation into condensations). To check convergence, we have repeated three simulations with $n_{_{\rm SPH}}\!\simeq\! 2,400,000 \; \rm pc^{-3}$; the resulting collapse times, density- and velocity-profiles differ by $\la 1\%$. We therefore believe that $n_{_{\rm SPH}}\!\simeq\! 800,000 \; \rm pc^{-3}$ delivers sufficiently high resolution. We have also simulated the high aspect fiducial case ($A\OO\!=\!10$, $\mu\OO\!=\!10\,{\rm M}_{_\odot}\,{\rm pc}^{-1}$, $R\OO\!=\!0.1\,{\rm pc}$) using a different SPH code \citep[{\sc seren};][]{Hub11}, and obtain almost identical results, so we are confident in the fidelity of the simulations.

\section{Results}\label{SEC:RES}%

\begin{figure*}
\centering
\includegraphics[width = 0.49\linewidth]{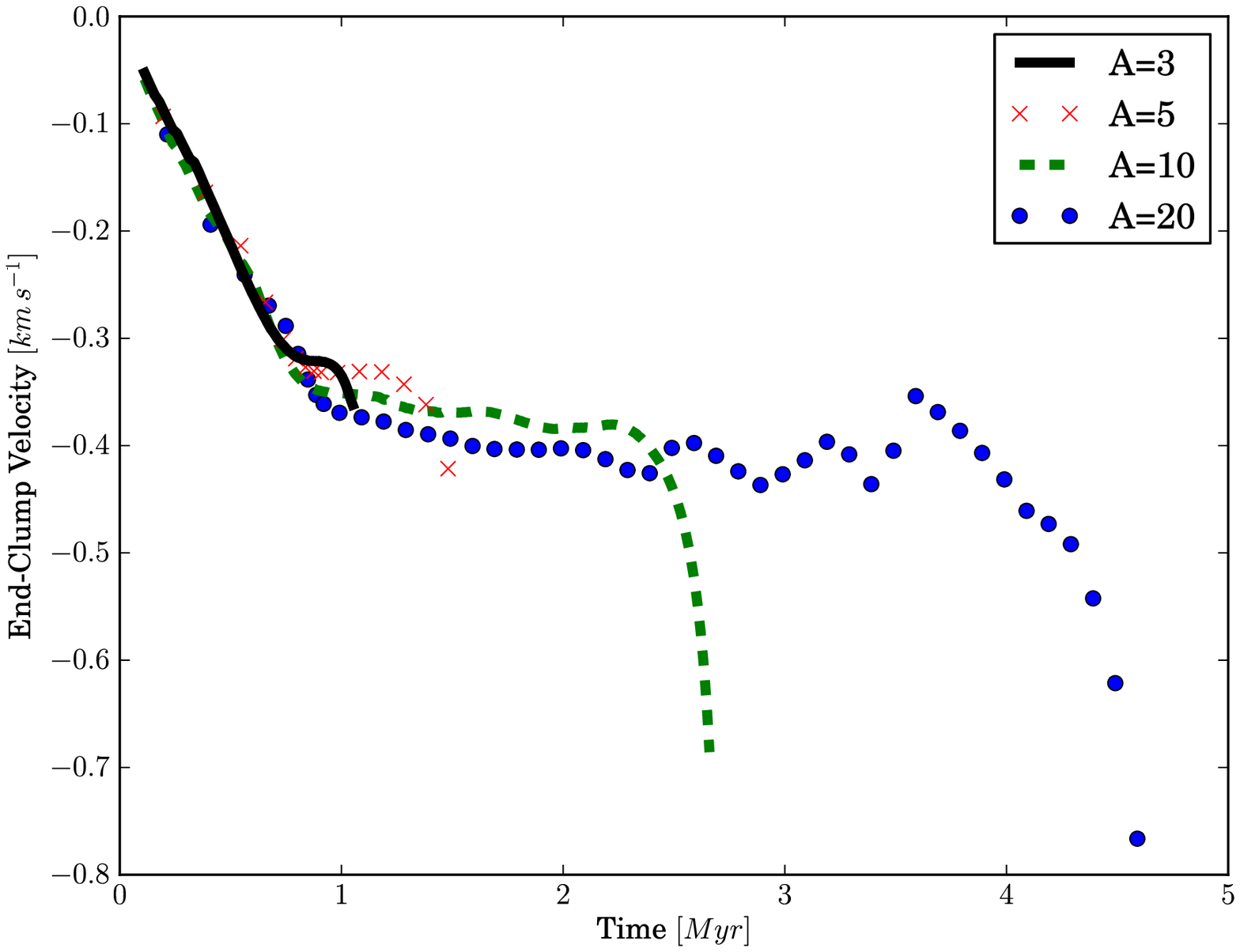}
\includegraphics[width = 0.49\linewidth]{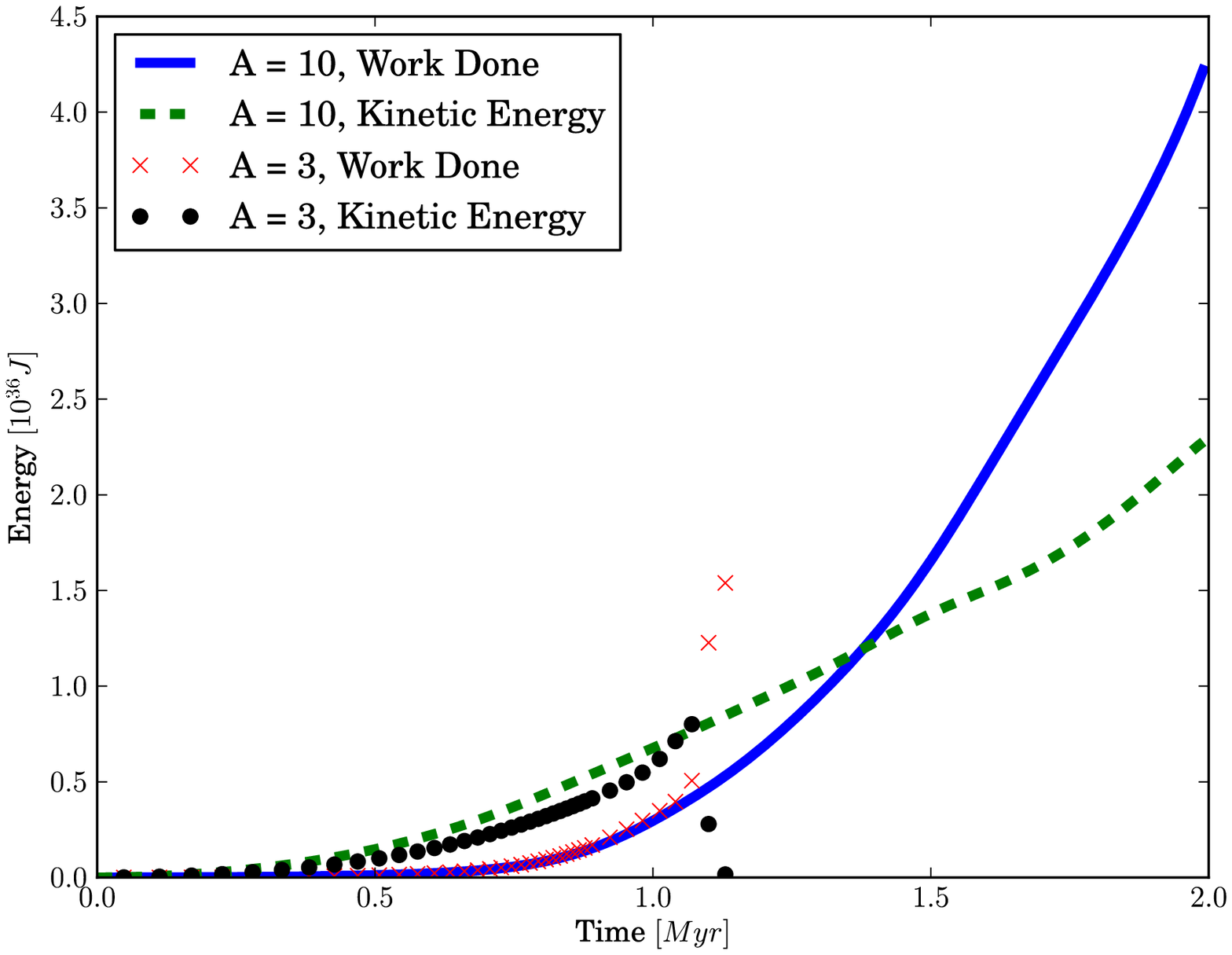}
\caption{(a) The velocity of the $z>0$ end-clump as a function of time for filaments with initial line-density $\mu\OO =10\,\rm M_{\odot}\,\rm{pc}^{-1}$, radius $R\OO =0.1\,\rm{pc}$, and different initial aspect ratios, $A\OO =3,\;5,\;10,\;{\rm and}\;20$. In the high aspect ratio simulations the end-clump accelerates inwards for the first $\sim 1\,{\rm Myr}$, but then approaches a terminal velocity, which is only weakly dependent on $A\OO$. The sudden increase in the negative velocity at the end of each profile is due to the two end-clumps becoming so close that the motion is dominated by their mutual gravitational attraction; there is essentially no gas left in between them to sweep up. (b) The total kinetic energy and the work done on the gas plotted against time for both the high and the low aspect ratio fiducial cases ($\mu\OO=10\,\rm M_{\odot}\,\rm pc^{-1}$, $R\OO=0.1\,\rm pc$ and $A\OO=10$ or $A\OO=3$). For the $A\OO = 10$ case, the end-clumps approach their terminal velocity, around $1\,{\rm Myr}$; thereafter, the rate at which work is done exceeds the rate of increase of the kinetic energy. This is not seen in the $A\OO = 3$ case due to the filament collapsing before terminal velocity is reached; the abrupt shift in energies at $ \sim \, 1.1 \,\rm Myr$ is due to the end-clumps meeting at $z=0$.}
\label{fig:energys}
\end{figure*}

For all aspect ratios, $2\la A\OO\la 20$, all line-densities, $2\,{\rm M}_{_\odot}\,{\rm pc}^{-1}\la\mu\OO\la 50\,{\rm M}_{_\odot}\,{\rm pc}^{-1}$, and all radii, $0.05\,{\rm pc}\la R\OO\la 0.15\,{\rm pc}$, the global collapse of the filament is end-dominated. The ends experience the highest acceleration, and this leads to the formation of end-clumps which then converge on the centre, sweeping up material as they go. Fig. \ref{fig:veldenprofile10} shows the density and velocity profiles along the $z$-axis, for both the high and low aspect ratio fiducial cases. The dense end-clumps and their supersonic inward velocities are clearly seen.

In all the simulations, we estimate the collapse time, $t_{_{\rm COL}}$, by identifying the moment when the two converging end-clumps merge, i.e. when there is a single central density maximum at $z=0$. The filled circles on Fig. \ref{fig:varythings}a show values of $t_{_{\rm COL}}$ obtained with different $A\OO$, when $\mu\OO$ and $R\OO$ are fixed at their fiducial values. The filled circles on Fig. \ref{fig:varythings}b show values of $t_{_{\rm COL}}$ obtained with different $\mu\OO$, when $A\OO$ and $R\OO$ are fixed at their high aspect ratio fiducial values. The filled circles on Fig. \ref{fig:varythings}c show values of $t_{_{\rm COL}}$ obtained with different $R\OO$, when $A\OO$ and $\mu\OO$ are fixed at their high aspect ratio fiducial values. The blue continuous curves show that all these results are well fitted with
\begin{equation}
\label{eq:collapsetime}
t_{_{\rm COL}} = (0.49 + 0.26A\OO)\,\left(G\rho\OO\right)^{-1/2}\,,
\end{equation}
where $\rho\OO\!=\!\mu\OO/\pi R\OO^2$ --- and hence they are not well fitted with the relations derived by \citet{Pon12} (i.e. Eqn. \ref{eq:homocollapse} for $A\OO\la 5$, and Eqn. \ref {eq:edgecollapse} for $A\OO\ga 5$).

Figure \ref{fig:predict} compares the collapse times obtained from simulations with $R\OO\!=\!0.1\,{\rm pc}$, $\mu\OO\!=\!35\,{\rm M}_{_\odot}\,{\rm pc}^{-1}$ or $\mu\OO\!=\!50\,{\rm M}_{_\odot}\,{\rm pc}^{-1}$, and a range of values for $A\OO$ (filled circles) with the predictions of Eqn. (\ref{eq:collapsetime}; continuous green and blue lines). Again the agreement is very good.

\section{Discussion}\label{SEC:DIS}%

The gravitational acceleration felt by particles off the symmetry axis is lower than that felt by those on the axis, an effect we discuss below in \S \ref{SUBSEC:SEMI}. This results in a curved end-clump, the portion of the end-clump on the axis being roughly $ 0.01 \; \rm pc$ ahead of the end-clump off axis. To determine the velocity of the end-clump, we locate the highest-density SPH particle that is at least $0.04 \, \rm pc$ off axis, and then average the velocity of all the SPH particles within $z \, \pm 0.02\,{\rm pc}$ of this particle. By selecting a SPH particle off axis we are able to avoid the atypical central region and locate the end-clump more accurately. The resulting velocity is insensitive to the precise choice of $\pm\, 0.02\,{\rm pc}$ and of $0.04 \, \rm pc$ as the off axis requirement.

Fig. \ref{fig:energys}a shows the velocity of the $z > 0$ end-clump as a function of time, for the four filaments with initial line-density $\mu\OO\!=\!10\,{\rm M}_{_\odot}\,{\rm pc}^{-1}$, radius $R\OO\!=\!0.1\,{\rm pc}$ and initial aspect ratios, $A\OO =3,\;5,\;10,\;{\rm and}\;20$. The velocity profiles are roughly independent of aspect ratio. For large values of $A\OO$ a terminal velocity is reached before the filament has time to fully collapse.

\begin{figure}
\centering
\includegraphics[width = 0.98\linewidth]{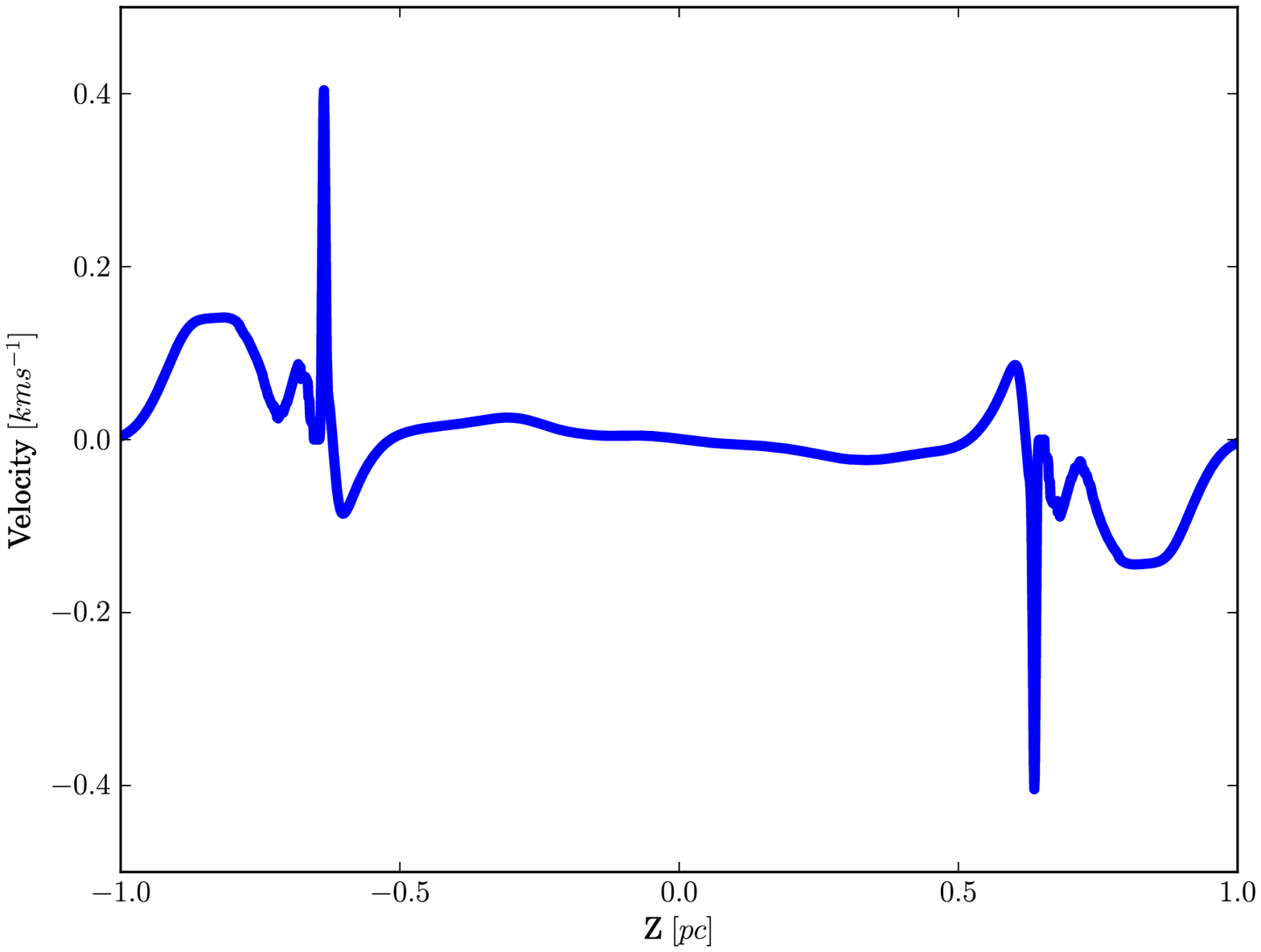}
\includegraphics[width = 0.98\linewidth]{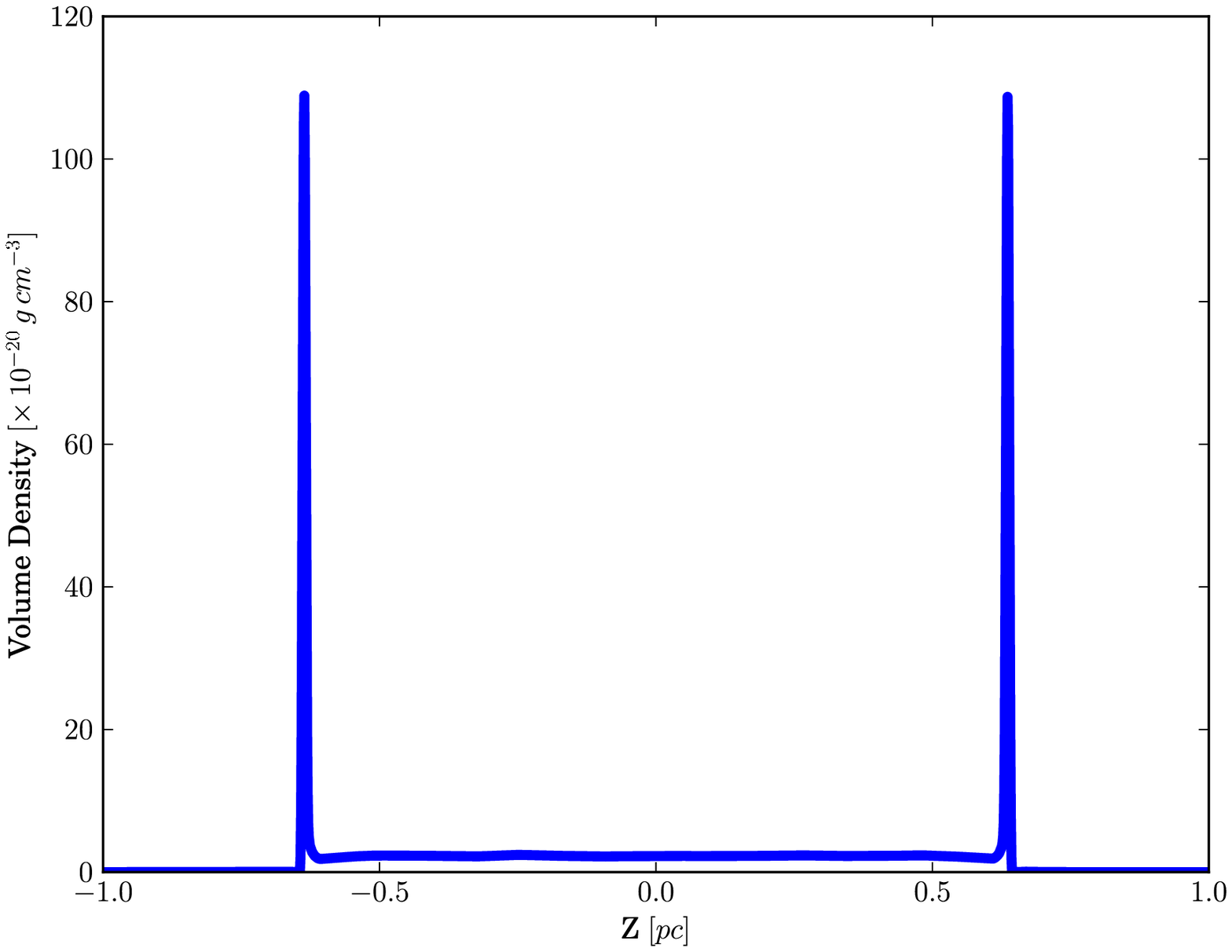}
\caption{(a) The gas velocity along the $z$-axis at time $t=1.2\,\rm{Myr}$, for the filament with $A\OO=10$, $\mu\OO=10\,\rm M_{\odot}\,\rm pc^{-1}$ and $R\OO=0.1\,\rm pc$. The gas just interior to the converging end-clumps is moving outwards, thereby increasing the ram-pressure that this gas delivers to the end-clump when it is swept up. (b) The density profile of the same filament at the same point in time. The density peaks (i.e the end-clumps) are at the same position as the velocity peaks. Note the uniformity of the interior density, it has remained roughly constant and uniform since the start of the simulation.}
\label{fig:A10vellater}
\end{figure}

A terminal velocity arises because, once an end-clump becomes sufficiently massive, it starts to dominate the gravitational field experienced by the gas just ahead of it, and so this gas flows outwards to meet the end-clump (see Fig. \ref{fig:A10vellater}a). This means that more work has to be done reversing the velocity of the swept-up gas, {\it and} the swept-up gas is more strongly compressed, so more work is done on it. The upshot is that less gravitational potential energy gets converted into kinetic energy, basically just enough to accelerate the newly swept-up gas up to the terminal velocity. This is illustrated on Fig. \ref{fig:energys}b, where we plot, for the simulations with the high and low aspect ratio fiducial parameters ( $R\OO=0.1\,\rm pc$, $\mu\OO=10\,\rm M_{\odot}\,\rm pc^{-1}$ and $A\OO=3$ or $A\OO=10$), the evolution of the net kinetic energy and the net work done. At early times, the kinetic energy grows fastest, but once the terminal velocity has been reached for the $A\OO=10$ case, around $\sim 1\,{\rm Myr}$, work done takes over, and the kinetic energy increases roughly linearly with time, due to the roughly linear increase in the mass of the end-clumps as they sweep in at constant velocity. This large growth in work done is absent for the $A\OO=3$ case; this is to be expected because the end-clumps merge in the centre before terminal velocity is reached. However, the same dynamics are present, the end-clump gains enough mass to dominate the gravitational field in its vicinity and the gas just ahead of it begins to move outwards to meet it.

\subsection{Semi-Analytical Model}\label{SUBSEC:SEMI}%

As well as considering the system in terms of energy, we can also develop a semi-analytical model on the basis of the forces involved. A terminal velocity is reached, for high aspect ratio filaments, because the gravitational force on the end-clump is approximately balanced by the force of ram-pressure due to the end-clump sweeping up the outward moving interior material. The net acceleration of the end-clump is therefore
\begin{equation}\label{EQN:ACCEL1}
\frac{d^2z}{dt^2}\;\,=\;\,-\,\alpha\,(2\,\pi\,G\,\rho\OO\,R\OO)\,+\,\frac{\rho\OO\,\Delta v_{_{\rm REL}}^{2}\,\pi\,R\OO^2}{M(t)}.
\end{equation}

The first term on the righthand side of Eqn. (\ref{EQN:ACCEL1}) is the gravitational acceleration that the end-clump experiences due to the interior material. $(2\pi G\rho\OO R\OO)$ is the acceleration derived by \citet{Pon12} from consideration of a point on the symmetry axis of the filament. $\alpha$ is a correction factor that takes account of the fact that the end-clump extends from the symmetry axis, $r\!=\!0$, to $r\!=\!R\OO$, where the gravitational acceleration is lower.

\begin{figure}
\centering
\includegraphics[width = 0.98\linewidth]{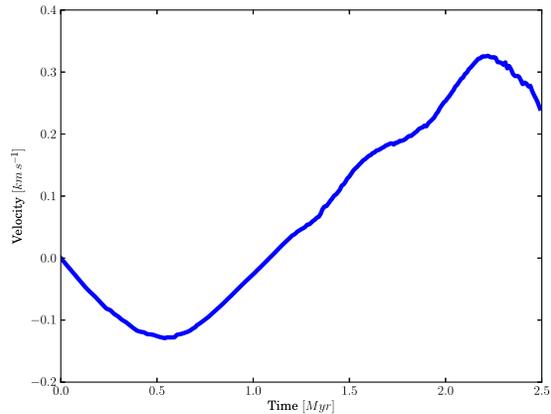}
\caption{The $z$-velocity of the gas just interior to the $z>0$ end-clump as a function of time. At early times, this gas flows inward with the end-clump, but at later times, it is attracted gravitationally to the end-clump and moves outwards to meet it. Close to the end of the collapse, the interior material begins to be gravitationally attracted by the $z<0$ end-clump resulting in the velocity maximum at $\sim \, 2.2 \, \rm Myr$.}
\label{fig:IntVel}
\end{figure}

The second term on the righthand side of Eqn. (\ref{EQN:ACCEL1}) is the acceleration experienced by the end-clump due to ram pressure; $\Delta v_{_{\rm REL}}$ is the relative velocity between the end-clump and the material about to be swept up, and $M(t)$ is the mass of the end-clump at time $t$. If the material interior to the end-clump were stationary, we could put $\Delta v_{_{\rm REL}}\!=\!dz/dt$, where $dz/dt$ is the velocity of the end-clump at time $t$. However, the interior material is not stationary, as seen in figure \ref{fig:A10vellater}a, and so we must put $\Delta v_{_{\rm REL}}\!=\!dz/dt-dz_{_{\rm I}}/dt$, where $dz_{_{\rm I}}/dt$ is the velocity of the material just interior to the end-clump at time $t$. Equation \ref{EQN:ACCEL1} is equivalent to the momentum equation used in the \citet{Pon12} paper. It is their assumption that $\Delta v_{_{\rm REL}}\!=\!dz/dt$ (i.e the interior material is stationary) which distinguishes their analysis from ours. 

Fig. \ref{fig:IntVel} shows that the velocity of the interior material is relatively low, and it moves inwards and then outwards. If one neglects the net motion of the interior material, then the mass of the end-clump is $M(t)\!\simeq\!\mu\OO(Z\OO-z(t))$, and the acceleration of the end-clump becomes 
\begin{equation}\label{EQN:ACCEL2}
\frac{d^2z}{dt^2}\;\,\simeq\;\,-\,\alpha\,(2\,\pi\,G\,\rho\OO\,R\OO)\,+\,\frac{(dz/dt - dz_{_{\rm I}}/dt)^{2} }{(Z\OO-z(t))}.
\end{equation}

\begin{figure}
\centering
\includegraphics[width = 0.98\linewidth]{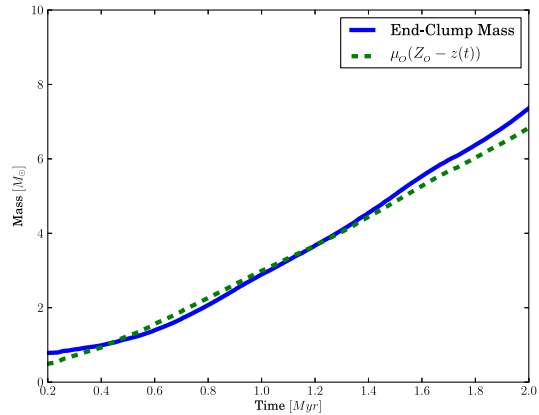}
\caption{The mass of the $z>0$ simulated end-clump is compared with the end-clump mass expected if the interior material was stationary, $M(t)\!=\!\mu\OO(Z\OO-z(t))$. The masses are within $\sim 10\%$ over the course of the simulation.}
\label{fig:ClumpMass}
\end{figure}

The assumption that the end clump's mass is approximately the same whether the interior material moves or not is shown to be valid in Fig. \ref{fig:ClumpMass}. Here, the mass of the $z > 0$ simulated end clump is compared with the function $M(t)\!=\!\mu\OO(Z\OO-z(t))$ for the high aspect fiducial case. The masses agree with each other within $ \sim 10 \%$. This is due to the interior density staying roughly constant during the collapse. At such low temperatures the end-clumps are moving supersonically and thus no density waves disturb the material interior to the clumps. Moreover, most of the interior material is unaffected by the gravitational pull of the end clumps and is approximately stationary until an end-clump approaches (Fig. \ref{fig:A10vellater}).  

To solve Eqn. (\ref{EQN:ACCEL2}) for $z(t)$ (the location of the end-clump), one must first find $dz_{_{\rm I}}/dt$, the speed of the material about to be swept up by the end-clump. The acceleration at $z_{_{\rm I}}$ is
\begin{eqnarray}\nonumber
\frac{d^2z_{_{\rm I}}}{dt^2}\!&\!\simeq\!&\!-\,\alpha(2\pi G\rho\OO)\left\{2z_{_{\rm I}}(t)+\left(R\OO^2+(z(t)\!-\!z_{_{\rm I}}(t))^2\right)^{1/2}\right.\\\nonumber
&&\hspace{2.85cm}\left.-\left(R\OO^2+(z(t)\!+\!z_{_{\rm I}}(t))^2\right)^{1/2}\right\}\\\nonumber
&&+\,\beta(2\pi G\rho\OO)(Z\OO\!-\!z(t))\left\{\frac{(z(t)\!+\!z_{_{\rm I}}(t))}{({R\OO^2\!+\!(z(t)\!+\!z_{_{\rm I}}(t))^{2}})^{1/2}}\right. \\\nonumber
&&\hspace{3.1cm}\left.-\frac{(z(t)\!-\!z_{_{\rm I}}(t))}{({R\OO^2\!+\!(z(t)\!-\!z_{_{\rm I}}(t))^{2}})^{1/2}}\right\}\\\label{EQN:ACCINT}
\end{eqnarray}

The first term on the righthand side of Eqn. (\ref{EQN:ACCINT}; the one preceded by $\alpha$) is the contribution to the gravitational acceleration from the material between the end-clumps, i.e. the acceleration at a point $z_{_{\rm I}}(t)$ in a uniform-density cylinder of radius $R\OO$ between $\pm z(t)$, where $z_{_{\rm I}}(t)\!<\!z(t)$ \citep{BurHar04}. We have again neglected motion of this material, and included the factor $\alpha$ to take account of the fact that the material away from the symmetry axis experiences a lower acceleration than the material on the axis.

The second term on the righthand side of Eqn. (\ref{EQN:ACCINT}; the one preceded by $\beta$) is the contribution to the gravitational acceleration from the end-clumps. This term is obtained by treating the end-clumps as discs of uniform surface-density, considering a point on the symmetry axis at $z_{_{\rm I}}(t)$, and then including a factor $\beta$ to allow for the fact that material away from the symmetry axis experiences a lower acceleration than the material on the axis.

Eqns. (\ref{EQN:ACCEL2}) and (\ref{EQN:ACCINT}) are second order non-linear coupled differential equations. We have solved them numerically for the high aspect ratio fiducial case ($A\OO\!=\!10$, $\mu\OO\!=\!10\,{\rm M}_{_\odot}\,{\rm pc}^{-1}$, $R\OO\!=\!0.1\,{\rm pc}$), using the first order Euler-Cromer method and $\delta t\!=\!0.003,\,0.010,\;{\rm and}\;0.020\,{\rm Myr}$; the results obtained with these different $\delta t$ values are essentially indistinguishable, indicating that the Euler-Cromer method is converged. The interior of the filament is divided into 500 evenly spaced Lagrangian discs and the velocity of each disc calculated at each time-step by integrating Eqn. (\ref{EQN:ACCINT}). The velocity of the disc just interior to the end-clump is then used to solve Eqn. (\ref{EQN:ACCEL2}). The correction factor $\alpha$ can be evaluated from the simulation data by averaging over the SPH particles in the end-clump, to obtain $\alpha\!=\!0.60$. The correction factor $\beta$ is obtained by averaging over the SPH particles making up discs of gas interior to the end-clumps, the first term of equation \ref{EQN:ACCINT} is then subtracted. This results in a value of $\beta\!=\!0.4$.

\begin{figure*}
\centering
\includegraphics[width = 0.49\linewidth]{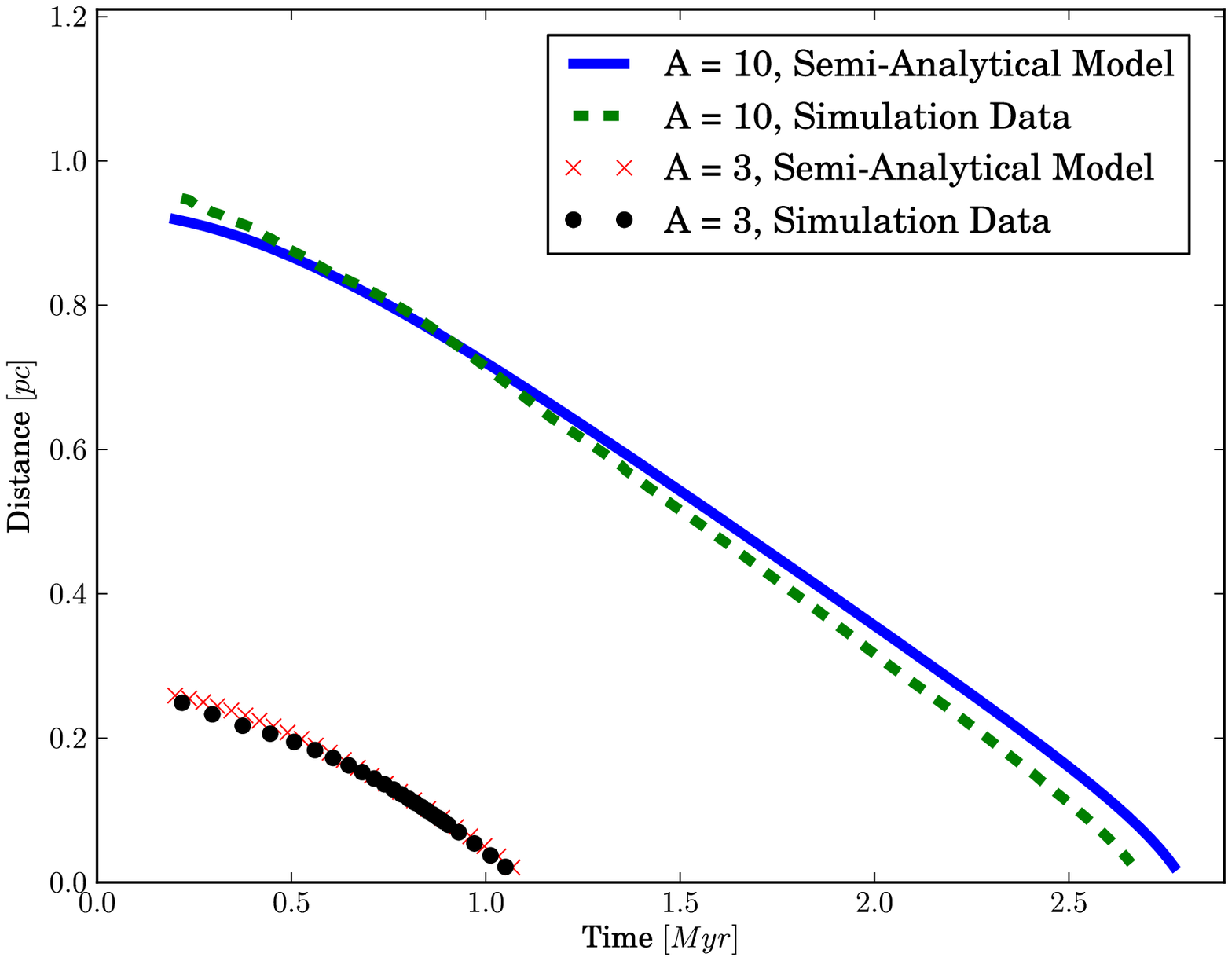}
\includegraphics[width = 0.49\linewidth]{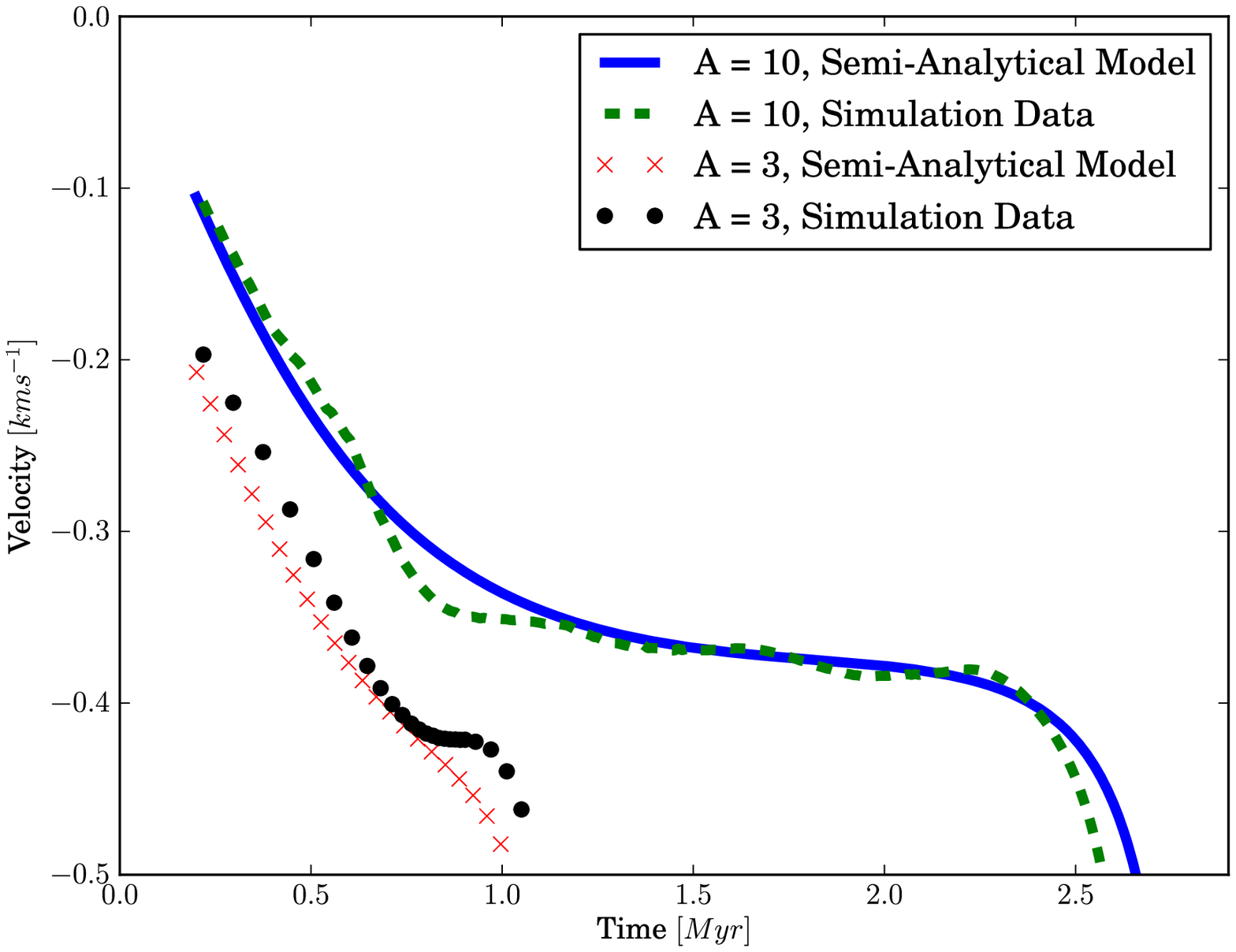}
\caption{ A comparison between simulation data and our semi-analytic model for a filament characterised by $\mu\OO\!=\!10\,{\rm M}_{_\odot}\,{\rm pc}^{-1}$, $R\OO\!=\!0.1\,{\rm pc}$ and either $A\OO\!=\!10$ or $A\OO\!=\!3$. (a) The position of the end-clump as a function of time. It starts at $t=0.2 \; \rm Myr$ when the end-clumps are clearly established. (b) The velocity of the end-clump as a function of time. The velocity of $A\OO =3$ end clump has been shifted downwards by $0.1 \; \rm km \, \rm s^{-1}$ for clarity. Simulation and model profiles fit reasonably well for both initial aspect ratios. }
\label{fig:modelvsim}
\end{figure*}

Fig. \ref{fig:modelvsim} compares the results of the semi-analytical model with the simulation, for the high aspect ratio fiducial case ($A\OO\!=\!10$, $\mu\OO\!=\!10\,{\rm M}_{_\odot}\,{\rm pc}^{-1}$, $R\OO\!=\!0.1\,{\rm pc}$) and the low aspect ratio fiducial case ($A\OO\!=\!3$, $\mu\OO\!=\!10\,{\rm M}_{_\odot}\,{\rm pc}^{-1}$, $R\OO\!=\!0.1\,{\rm pc}$); we have also made this comparison for the cases with $A\OO\!=\!10$, $R\OO\!=\!0.1\,{\rm pc}$, and $\mu\OO\!=\!35\,{\rm M}_{_\odot}\,{\rm pc}^{-1}\;{\rm and}\;50\,{\rm M}_{_\odot}\,{\rm pc}^{-1}$, and the agreement is equally good. This suggests that the semi-analytical model is capturing the important physical processes, and that the factors $\alpha\!=\!0.60$ and $\beta\!=\!0.4$ are valid correction terms.

\begin{figure}
\centering
\includegraphics[width = 0.98\linewidth]{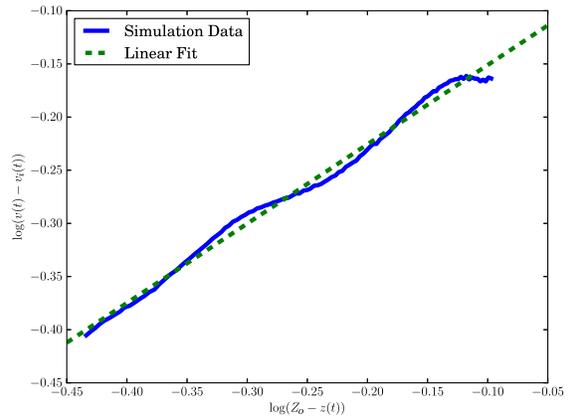}
\caption{A test of the validity of equation \ref{eq:AtTerminal}; $\log_{_{10}}(dz/dt\,-\,dz_{_{\rm I}}/dt)$ is plotted against $\log_{_{10}}(Z\OO\,-\,z(t))$ for the high aspect ratio fiducial case. The simulation data are well fit by a straight line with a gradient of 0.64, close to the expected 0.5. The oscillations around the linear fit show that once the terminal velocity is approached, the dynamics is stable.}
\label{fig:AtTerminal}
\end{figure}

If Eqn. (\ref{EQN:ACCEL2}) accurately describes the acceleration of an end-clump, then, once the end-clump approaches its terminal velocity and its acceleration falls towards zero, we must have
\begin{equation}
\label{eq:AtTerminal}
\frac{(dz/dt\,-\,dz_{_{\rm I}}/dt)^{2} }{(Z\OO\,-\,z(t))}\;\;\sim\;\;\alpha\,2\,\pi\,G\,\rho\OO\,R\OO\,.
\end{equation} 
Since the righthand side of this equation is constant, a plot of $\log_{_{10}}(dz/dt\,-\,dz_{_{\rm I}}/dt)$ against $\log_{_{10}}(Z\OO\,-\,z(t))$ should have slope $0.5$. This is shown in figure \ref{fig:AtTerminal} for the high aspect ratio fiducial case ($A\OO\!=\!10$, $\mu\OO\!=\!10\,{\rm M}_{_\odot}\,{\rm pc}^{-1}$, $R\OO\!=\!0.1\,{\rm pc}$) between $1.2\,{\rm Myr}$ and $2.2\,{\rm Myr}$. It is well fit by a straight line with a gradient of 0.64, close to the expected 0.5. The deviation is due to the fact that the end-clump is not quite moving at a constant velocity; it experiences a small acceleration. There is a small oscillation about the straight line, indicating that, once the terminal velocity is approached, the dynamics is stable: if the end-clump is moving slightly slower than the terminal velocity, the acceleration due to ram pressure is slightly less than the gravitational acceleration, and it speeds up; conversely, if the end-clump is moving slightly faster than the terminal velocity, the acceleration due to ram-pressure is slightly greater than the gravitational acceleration, and it slows down. 

Eqn. (\ref{eq:collapsetime}) agrees with the generic theorem that the free-fall collapse time for a uniform-density filament is longer than for a sphere of the same uniform density. However, it disagrees with the predictions of \citet{Pon12} (Figure \ref{fig:varythings}a). Overall, \citet{Pon12} predicts somewhat longer collapse times for short filaments, $A\OO\!\la\!9$, and shorter collapse times for longer filaments, $A\OO\!\ga\!9$. The \citet{Pon12} collapse times for short filaments are too long, because their analysis does not take account of the fact that even in this case there are end-clumps. The \citet{Pon12} collapse times for long filaments are too short, because their analysis does not take account of the interaction between an end-clump and the material it is about to sweep up, as analysed in \S \ref{SEC:DIS}. We do not consider the discrepancies between our results and those found in \citet{Pon12} for the range $A\OO \, < \, 3$ as such a cloud is outside our definition of a filament.   

\section{Conclusions}\label{SEC:CON}%

\citet{Pon12} have suggested that there are two modes for the  freefall collapse of a uniform-density filament: short filaments, $A\OO\!\la\!5$, collapse homologously, on a timescale that varies as $A\OO$ (see Eqn. \ref{eq:homocollapse}); longer filaments undergo end-dominated collapse, on a timescale that varies as $A\OO^{\!1/2}$ (see Eqn. \ref{eq:edgecollapse}). We find no such dichotomy. Rather, end effects are evident for all aspect ratios, and just one equation, 
\begin{equation}
t_{_{\rm COL}} = (0.49 + 0.26A\OO)\,\left(G\rho\OO\right)^{-1/2} \tag{\ref{eq:collapsetime}}
\end{equation}
predicts the free-fall collapse times obtained in our simulations to within $\pm 3\%$ for $2\la A\OO\la 20$, $2\,{\rm M}_{_\odot}\,{\rm pc}^{-1}\la \mu\OO \la 50\, {\rm M}_{_\odot}\,{\rm pc}^{-1}$, and $0.05\,{\rm pc}\la R\OO\la 0.15\,{\rm pc}$.

The linear relationship between $t_{_{\rm COL}}$ and $A\OO$, for large aspect ratios, is due to the fact that the end-clumps quickly approach a terminal velocity before collapse finishes. The same linear relationship still holds for small aspect ratio filaments despite the lack of a terminal velocity. Hence, the linear relationship is ultimately a result of the dynamics of the system which are independent of initial aspect ratio. Before being swept up, the gas immediately ahead of an end-clump is accelerated outwards by the gravitational attraction of the approaching end-clump. This outward velocity increases the amount of ram pressure exerted on the end-clump, and the amount of work needed to accelerate and compress the swept up gas. \citet{Pon12} appear to have neglected this outward moving gas and its consequences.

We have constructed a semi-analytical model which describes the acceleration of the gas in a collapsing filament, using only gravity and ram-pressure. This model is able to reproduce the end-clump position and velocity seen in the simulations and accurately describes the evolution of the end-clumps over a range of initial conditions. 

\section{Acknowledgments}\label{SEC:ACK}%

SDC gratefully acknowledges the support of a STFC postgraduate studentship. APW gratefully acknowledges the support of a consolidated grant (ST/K00926/1) from the UK STFC. We would like to thank D. Hubber and G. Rosotti for the use of their new SPH code, GANDALF. We also thank the anonymous referee for their very useful comments which helped to improve this paper. This work was performed using the computational facilities of the Advanced Research Computing at Cardiff (ARCCA) Division, Cardiff University. 

\bibliographystyle{mn2e}
\bibliography{ref} 

\label{lastpage}

\end{document}